\documentclass[pra,10pt,notitlepage,superscriptaddress,tightenlines,amsmath,amssymb,showpacs,eqsecnum]{revtex4-1}

\usepackage{bm,graphicx,hyperref,upgreek,empheq,color}
\usepackage[normalem]{ulem}

\newcommand{\braket}[2]{\langle #1 | #2 \rangle}

\def\UpsilonT{\Upsilon_{\!T}}

\begin{document}

\title{Reduced dimensionality and spatial entanglement in highly anisotropic Bose-Einstein condensates}

\author{Alexandre B.~Tacla}

\affiliation{
Center for Quantum Information and Control, MSC07--4220, University of New Mexico,
Albuquerque, New Mexico 87131-0001, USA}

\author{Carlton M.~Caves}
\affiliation{
Center for Quantum Information and Control, MSC07--4220, University of New Mexico, Albuquerque, New Mexico 87131-0001, USA}
\affiliation{Centre for Engineered Quantum Systems, School of Mathematics and Physics, University of Queensland, Brisbane, Queensland 4072, Australia}

\begin{abstract}
We investigate the reduced dimensionality of highly anisotropic Bose-Einstein condensates (BECs) in connection with the entanglement between the spatial degrees of freedom.  We argue that the reduced dimensionality of the BEC is physically meaningful in a regime where spatial correlations are negligible.  We handle the problem analytically within the mean-field approximation for general quasi-one-dimensional and quasi-two-dimensional geometries and obtain the optimal reduced-dimension, pure-state description of the condensate mean field.  We give explicit solutions for the case of harmonic potentials, which we compare against exact numerical integration of the three-dimensional Gross-Pitaevskii equation. %
\end{abstract}

\maketitle

% ================================================================================
\section{Introduction \label{sec:introduction}} % (fold)

Bose-Einstein condensates (BECs) confined in highly anisotropic potentials are commonly used to explore physical phenomena particular to lower-dimensional geometries, such as the Berezinsky-Kosterlitz-Thouless phase transition~\cite{hadzibabic}, as well as to implement technological devices, such as atom lasers~\cite{robins} and metrological sensors~\cite{wildermuth,boixo09}.  Thanks to high levels of experimental control, the reduced-dimension regime of highly anisotropic condensates can be achieved by making the characteristic energy scale of the tightly confined (transverse) dimension(s) much higher than the interaction energy of the atomic cloud~\cite{gorlitz01}.  When the separation in energy scales is sufficiently large, nontrivial transverse spatial modes become virtually inaccessible and the system exhibits lower-dimensional behavior.  In this regime, the transverse spatial co\"ordinates become physically redundant and can be effectively eliminated from the description of the gas in favor of a simpler reduced-dimension model, which involves only the lower-energy, longitudinal degrees of freedom.

Within the mean-field approximation, dimensional reduction of the three-dimensional (3D) Gross-Pitaevskii (GP) equation is often derived under the assumption that, due to the tight confinement, the scattering interaction does not couple to the high-energy transverse degrees of freedom~\cite{pitaevskii}.  Given this assumption, the condensate mean field is factorizable relative to the longitudinal and transverse co\"ordinates, with the latter being described by the ground-state wave function of the bare transverse trapping potential.  Theoretical analyses and recent experiments have shown, however, that in spite of the high anisotropy of the gas, such an idealized picture is not generally valid.  Effects of the scattering interaction on the transverse spatial modes have been shown to impact key properties of the gas, such as density propagation~\cite{muryshev02,sinha06,khaykovich06}, even beyond mean field, such as thermalization~\cite{mazets,gring}, and have been predicted to produce significant deviations in interferometry of quasi-1D BECs~\cite{tacla10,tacla13}.

Variational \emph{Ans\"atze} are often proposed to derive effective reduced-dimension models that capture 3D-induced effects~\cite{salasnich,das02a,das02b,gerbier04,munozmateo}. Although sometimes effective, variational methods rely on various \emph{a priori\/} assumptions about the condensate mean field and \emph{ad hoc\/} trial functions that cannot be applied to general trapping potentials.  In this paper, we instead use the perturbative method developed in~\cite{tacla11} to study the reduced dimensionality of highly anisotropic BECs and its connection to the entanglement between the condensate's spatial degrees of freedom.  We show that this technique provides an \emph{unbiased\/} way to characterize the reduced dimensionality of the condensate as a regime of small spatial entanglement for general quasi-1D and quasi-2D geometries.  Moreover, this formalism provides the optimal product-state approximation to the exact 3D condensate wave function.  This approximation, valid within the regime of low spatial correlations, decouples the transverse and longitudinal spatial degrees of freedom; we show that regardless of the trapping geometry, the resulting reduced-dimension mean field corresponds to the solution of a longitudinal GP equation with additional attractive, three-body interactions.

By applying a Thomas-Fermi approximation to the longitudinal GP equation, we determine the reduced-dimension mean field analytically and derive a full 3D analytical estimate of the condensate's ground-state wave function.  We use this analytical wave function to study the role of spatial correlations in highly anisotropic BECs and to derive formulas for various ground-state properties of the condensate, such as the chemical potential and average density. Such estimates are valid as long as the nonlinear scattering energy is much smaller than the transverse energy scales and much larger than the longitudinal kinetic energy.  We assess the validity and performance of  the (analytical) optimal product-state approximation by direct comparison against numerical integration of the 3D GP equation for quasi-1D and -2D potentials with different aspect ratios and various atom numbers. In addition, we benchmark our approach against the commonly adopted variational \emph{Ans\"atze} of~\cite{salasnich}, which assume a Gaussian distribution with variable width along the transverse dimension(s).  Although our method applies to general trapping potentials, in this paper we focus on the case of quasi-1D and -2D harmonic traps, for which the variational technique~\cite{salasnich} is also applicable.

% section introduction (end)

% ================================================================================
\section{Reduced-dimension model \label{sec:reduced_dimension_model}} % (fold)

Highly anisotropic BECs are typically trapped by a loose longitudinal potential $V_L({\bm r})$ in $d$ dimensions, while the remaining $D=3-d$ transverse spatial degrees of freedom are tightly confined by a potential $V_T({\bm\rho})$.  Within the mean-field approximation, the condensate's ground-state wave function (normalized to unity) is defined by the 3D GP equation
\begin{equation}
\label{3DGPE}
\mu\uppsi({\bm \rho},{\bm r}) = \Big( H_{T} +  \epsilon H_{L} + \epsilon\tilde g|\uppsi({\bm \rho},{\bm r})|^2 \Big)\uppsi({\bm \rho},{\bm r}).
\end{equation}
Here $H_{T(L)}=-(\hbar^2/2M)\nabla_{T(L)}^2 + V_{T(L)}$ is the transverse (longitudinal) single-particle Hamiltonian, $\mu$ is the chemical potential, and $\tilde g = (N-1)g$, where $g=4\pi\hbar^2a/M$ is the scattering strength determined by the $s$-wave scattering length $a$, the atomic mass $M$, and the number of condensed atoms $N$.  In Eq.~(\ref{3DGPE}) we introduce a formal perturbation parameter $\epsilon$ to make explicit the physical difference between the high-energy scale of the tightly confined dimensions and the much lower energy scale of the longitudinal degrees of freedom and of the scattering interaction.  This perturbative regime holds as long as the number of atoms in the condensate is small compared to an (upper) critical atom number $N_T$.  The perturbation parameter thus characterizes the reduced-dimension regime and should be set equal to 1 at the end of the calculation. We discuss the physical dimensionless expansion parameter and the critical atom number $N_T$ in Appendix~\ref{ap:critical_atom_numbers}.

Because of the nonlinear scattering interaction, solutions to Eq.~(\ref{3DGPE}) are not spatially separable for nonuniform trapping potentials.  Nevertheless, there exists a \emph{unique\/} pair of orthonormal basis sets $\{\upchi_{n}({\bm \rho})\}$ and $\{\upphi_{n}({\bm r})\}$, which allows the condensate mean-field solution to be written in the form
\begin{align}
\label{Schmidt}
	\uppsi({\bm \rho},{\bm r}) = \sum_{n=0}^{\infty} \sqrt{\uplambda_n} \upchi_n({\bm \rho}) \upphi_n({\bm r}).
\end{align}
Equation~(\ref{Schmidt}) is known as the Schmidt decomposition and fully characterizes the entanglement between the transverse and longitudinal degrees of freedom~\cite{nielsen}. Finding the Schmidt decomposition~(\ref{Schmidt}) is, of course, equivalent to solving Eq.~(\ref{3DGPE}).  Formally, it requires diagonalizing the reduced transverse and longitudinal density matrices,
\begin{align}
	n_T({\bm \rho},{\bm \rho'}) &= \int d^d r\,\uppsi({\bm \rho},{\bm r})\uppsi^*({\bm \rho'},{\bm r}),\label{nT}\\
	n_L({\bm r},{\bm r'}) &= \int d^D\!\rho\,\uppsi({\bm \rho},{\bm r})\uppsi^*({\bm \rho},{\bm r'}),\label{nL}
\end{align}
which have $\upchi_{n}({\bm \rho})$ and $\upphi_{n}({\bm r})$ as their respective eigenfunctions and $\uplambda_n$ as their (nonnegative) common eigenvalues.  This approach, however, requires prior knowledge of the condensate mean field and, hence, is not effective in determining the Schmidt basis when $\uppsi({\bm \rho},{\bm r})$ is unknown.

In the reduced-dimension regime, however, we found in~\cite{tacla11} a way to simplify this problem dramatically; our method solves the 3D GP equation perturbatively by requiring its solution to be in Schmidt form.  The simplification arises from the fact that when the nonlinear interaction is small compared to the transverse energy scales, the entanglement between ${\bm \rho}$ and ${\bm r}$ is also small, appearing as higher-order corrections in $\epsilon$. In other words, the perturbation theory assumes that the solution to Eq.~(\ref{3DGPE}) in the reduced-dimension regime is close to a product state and, for this reason, that the Schmidt decomposition of the 3D condensate wave function $\uppsi({\bm \rho},{\bm r})$ has only two terms to first-order in $\epsilon$, i.e.,
\begin{align}
\label{psiSchmidt}
  \uppsi({\bm \rho},{\bm r}) \simeq \sqrt{\lambda_0} \chi_0({\bm \rho};\epsilon) \phi_0({\bm r};\epsilon) + \epsilon \sqrt{\lambda_1} \chi_1({\bm \rho};\epsilon) \phi_1({\bm r};\epsilon),
\end{align}
where the second term on the right-hand side is treated formally as a perturbation to the first one.  Note that the solution to Eq.~(\ref{3DGPE}) is spatially separable (i.e., has a single term in the Schmidt decomposition) when $\epsilon\rightarrow 0$.  This method is developed in detail in~\cite{tacla11} and generalized to the time-dependent case as well as to two-mode BECs in~\cite{tacla13}.

We summarize now the main results of~\cite{tacla11}, which we use throughout this paper. For simplicity, we assume from now on that the condensate wave function and the Schmidt functions are real. It is also convenient to absorb the Schmidt coefficients into the transverse Schmidt basis functions, which thus satisfy the relation
\begin{equation}\label{chinchim}
\langle\chi_n|\chi_m\rangle=\lambda_n\delta_{nm},
\end{equation}
where the bra-ket notation is used here and henceforth to denote spatial integrals.

The first (and dominant) Schmidt term is given by
\begin{align}
\chi_{0}({\bm \rho})&=\xi_0({\bm \rho})-\epsilon\tilde g\eta_L\sum_{n=1}^\infty \frac{\langle\xi_n|\xi_0^3\rangle}{E_n-E_0}\xi_n({\bm \rho}),\label{chi0}\\
\mu_L \phi_0({\bm r})
    &= \big[H_L + \tilde g \eta_{T} \phi_0^2({\bm r}) - 3\epsilon\tilde g^2\UpsilonT\phi_{0}^4({\bm r})\big]\phi_0({\bm r}),\label{phi0}
\end{align}
where Eq.~(\ref{chi0}) is written in terms of the eigenfunctions and eigenenergies of the bare transverse Hamiltonian $H_T$ (i.e., $H_T\xi_n=E_n\xi_n$) and $\mu_L=\mu-E_0$ is the longitudinal part of the chemical potential.  This solution assumes that $H_T$ has a nondegenerate ground state. In the limit $\epsilon\rightarrow 0$, the scattering interaction does not couple to the transverse degrees of freedom; in this limit, $\chi_0\rightarrow \xi_0$, $\phi_0\rightarrow\phi_{00}$, $\mu_L\rightarrow\mu_{L0}$, and Eq.~(\ref{phi0}) reduces to
\begin{align}
\mu_{L0} \phi_{00} &= \big(H_L + \tilde g \eta_{T} \phi_{00}^2 \big)\phi_{00}.\label{phi00}
\end{align}
The quantities
\begin{align}
	\eta_T&=\braket{\xi_0}{\xi_0^3},\label{etaT}\\
	\eta_L&=\braket{\phi_{00}}{\phi_{00}^3},\label{etaL} \\
	\UpsilonT&=
\sum_{n=1}^\infty\frac{\langle\xi_n|\xi_0^3\rangle^2}{E_n-E_0}\ge0,\label{UpsilonT}
\end{align}
renormalize the interaction strengths according to the geometry of the trapping potentials. Note that $\eta_{T(L)}$ is the average of the probability distribution $\xi_0^2({\bm \rho})$ [$\phi_{00}^2({\bm r})$] over itself.  Thus for a cigar-shaped condensate ($d=1$), $\eta_{T(L)}$ provides a measure of the inverse transverse (longitudinal) cross-sectional area (length) of the condensate, and for a pancake-shaped condensate ($d=2$), it provides a measure of the thickness (face area) of the condensate.

Consistent with the perturbation theory, Eq.~(\ref{chi0}) shows that $\chi_0$ is the ground-state wave function of the effective transverse potential $V_T({\bm \rho})+\epsilon\tilde g \eta_L \xi_0^2({\bm \rho})$ with eigenenergy equal to $E_0+\epsilon\tilde g \eta_L\eta_T$. The longitudinal wave function $\phi_0$, on the other hand, is the solution to a reduced-dimension, GP-like equation with an additional quintic term, which acts as an effective three-body, attractive interaction among the atoms.  These effective interactions are mediated by changes in the transverse wave function and can be interpreted as coming from a rectified $\eta_T$, given by
\begin{align}
	\tilde \eta_T({\bm r}) = \eta_T - 3 \epsilon\tilde g\UpsilonT\phi_0^2({\bm r}),
\end{align}
which is a function of the local longitudinal distribution $\phi_0^2({\bm r})$ and, hence, takes into account inhomogeneities in the trapping confinement.

The second Schmidt term is responsible for introducing entanglement between the transverse and longitudinal degrees of freedom and is given by
\begin{align}
	\chi_{1}({\bm \rho})&=-\tilde g\Delta\eta_L \sum_{n=1}^\infty \frac{\langle\xi_n|\xi_0^3\rangle}{E_n-E_0}\xi_n({\bm \rho}),\label{chi1}\\
	\phi_{1}({\bm r}) &= \frac{\phi_{00}^2({\bm r}) - \eta_{L}}{\Delta\eta_L}\phi_{00}({\bm r}),
\label{phi1}
\end{align}
where
\begin{align}
\Delta\eta_L^{2}\equiv\langle \phi_{00}^3 |\phi_{00}^3 \rangle - \eta_{L}^2\ge0
\end{align}
is the variance of $\phi_{00}^2$.  Note that in the case of a homogeneous longitudinal potential, $\Delta\eta_L=0$, and the solution to Eq.~(\ref{3DGPE}) is spatially separable.

The main objective of this paper is to use this formalism to assess the importance of the spatial correlations in the reduced-dimension regime of highly anisotropic condensates.  Although formally spatially entangled, we argue that the reduced dimensionality of the BEC is physically meaningful in a regime where spatial correlations are negligible.  Moreover, because the perturbation expansion is tied to the Schmidt decomposition, the dominant Schmidt term $\chi_0({\bm \rho})\phi_0({\bm r})$ is by construction the best product approximation to the exact condensate wave function $\uppsi({\bm \rho},{\bm r})$~\cite{lockhart02}. Indeed, note that to first order in $\epsilon$, $\chi_0$ and $\phi_0$ provide directly an estimate for the reduced density matrices~(\ref{nT}) and (\ref{nL})
\begin{align}
	n_T({\bm \rho},{\bm \rho'}) &=\chi_0({\bm \rho})\chi_0({\bm \rho'}) + O(\epsilon^2),\label{nTschmidt}\\
	n_L({\bm r},{\bm r'}) &=\phi_0({\bm r})\phi_0({\bm r'}) + O(\epsilon^2)\label{nLschmidt}.
\end{align}
As a result, the dominant longitudinal Schmidt function $\phi_0({\bm r})$ corresponds to the optimal reduced-dimension, pure-state description of the condensate mean field.  In addition, we determine below the reduced-dimension mean field $\phi_0({\bm r})$ within the longitudinal Thomas-Fermi (TF) approximation and thus obtain a fully analytical estimate of the mean field $\uppsi({\bm \rho},{\bm r})$.  This procedure allows one to derive accurate analytical formulas for various physical properties of the~BEC in the reduced-dimension regime.

% section reduced_dimension_model (end)

% ================================================================================
\section{Longitudinal Thomas-Fermi approximation} % (fold)
\label{sec:longitudinal_thomas_fermi_approximation}

The perturbation theory discussed above is valid for both repulsive and attractive condensates (provided the ground state is stable).  For repulsive BECs, however, one can easily solve Eqs.~(\ref{phi00}) and~(\ref{phi0}) in the longitudinal TF approximation by dropping the kinetic energy terms in those equations.  Naturally, this approximation requires the nonlinear interaction terms to be much larger than the neglected terms, which holds as long as the number of atoms in the condensate is large compared to a lower critical atom number $N_L$.  We discuss the validity of the longitudinal TF approximation and define $N_L$ in Appendix~\ref{ap:critical_atom_numbers}.

Within such an approximation, Eq.~(\ref{phi0}) assumes the simple form, quadratic in $\phi_0^2(\bm r)$,
\begin{align}
\label{effGPE_TFapprox}
	\mu_L-V_L({\bm r}) - \tilde{g}\eta_{T}\phi_0^2({\bm r}) + 3 \epsilon\tilde{g}^2\UpsilonT\phi_{0}^4({\bm r}) = 0,
\end{align}	
whose perturbative solution is given by
\begin{equation}
\label{phi0TF}
	\phi_0^2({\bm r}) = \frac{\mu_L-V_L({\bm r})}{\tilde{g}\eta_T} + \epsilon \frac{3\tilde{g}\UpsilonT}{\eta_T}\!\left( \frac{\mu_L-V_L({\bm r})}{\tilde{g}\eta_T} \right)^{\!2}.
\end{equation}
In the limit $\epsilon\rightarrow 0$, Eq.~(\ref{phi0TF}) reduces to the TF solution to Eq.~(\ref{phi00}), i.e.,
\begin{equation}
\label{phi00TF}
	\phi_{00}^2({\bm r}) = \frac{\mu_{L0}-V_L({\bm r})}{\tilde{g}\eta_T}.
\end{equation}

Equation~(\ref{phi0TF}), together with Eqs.~(\ref{chi0}), (\ref{chi1}), and (\ref{phi1}), forms the first-order analytical solution to the 3D GP equation, valid in the reduced-dimension regime, for arbitrary potentials of the form $V_T({\bm \rho})+V_L({\bm r})$.  The longitudinal chemical potentials $\mu_L$ and $\mu_{L0}$ are obtained by imposing the positivity and normalization constraints on Eqs.~(\ref{phi0TF}) and (\ref{phi00TF}).

For instance, for $d$-dimensional longitudinal, power-law potentials of the form
\begin{align}\label{powerlawVL}
	V_L({\bm r})=\frac12 k |{\bm r}|^{q},
\end{align}
we define
\begin{align}
	\mu_L= \frac{1}{2} k R_L^{q},\label{muL}
\end{align}
where $R_L$ is the radius at which $\phi_0^2(\bm r)$ goes to zero (i.e., the radius where $V_L=\mu_L$).  Normalization of $\phi_0^2(\bm r)$ for $r\le R_L$ then implies that
\begin{align}
	1 = \frac{k}{\tilde{g}\eta_T}\frac{q \pi^{d-1}}{d(q+d)}\!\left( 1 +\epsilon\frac{3\UpsilonT}{\eta_{T}^2} \frac{q}{2q+d} k R_L^{q}  \right) R_L^{q+d},\label{RL}
\end{align}
which defines $R_L$, the longitudinal TF length for $d=1$ or radius for $d=2$.  Note that the quantity $\eta_T^2/\UpsilonT$ provides the relevant quantification of the transverse energy scale as far as the perturbation theory is concerned.

Equation~(\ref{RL}) can be easily solved perturbatively by seeking a solution in the form
\begin{align}
	R_L = R_{L0} + \epsilon R_{L1}.\label{RL1st}
\end{align}
Here $R_{L0}$, found by setting the perturbation parameter $\epsilon$ to zero,
\begin{align}
	R_{L0}&= \left( \frac{\tilde{g}\eta_T}{k}\frac{d(q+d)}{q \pi^{d-1}}\right)^{1/(q+d)},\label{RL0}
\end{align}
is the radius at which $\phi_{00}^2(\bm r)$ goes to zero (i.e., the radius where $V_L=\mu_{L0}$); it determines the zero-order longitudinal chemical potential through
\begin{align}
\mu_{L0}= \frac{1}{2} k R_{L0}^{q}.\label{muL0}
\end{align}
It is easy to see that the first-order correction $R_{L1}$ is given by
\begin{equation}
	\frac{R_{L1}}{R_{L0}}
= - \frac{3\UpsilonT}{\eta_{T}^2} \frac{q}{(2q+d)(q+d)}k R_{L0}^q.\label{RL1}
\end{equation}
Plugging these results back into the expression~(\ref{muL}) for the longitudinal chemical potential gives, to first order in $\epsilon$,
\begin{align}
	\mu_L= \frac{1}{2} k R_{L0}^{q} \!\left( 1 + \epsilon q\frac{R_{L1}}{R_{L0}}\right)
	=\frac{1}{2} k R_{L0}^{q}
\!\left( 1 - \epsilon \frac{3\UpsilonT}{\eta_{T}^2}\frac{q^2}{(2q+d)(q+d)}kR_{L0}^q\right).\label{muL1st}
\end{align}

Note that for a harmonic transverse trapping potential, $V_T(\bm \rho)=\frac12 M\omega_T^2{\bm\rho}^2$, we have $\eta_T=1/(2\pi)^{D/2}\rho_0^D$, where $\rho_0=\sqrt{\hbar/M\omega_T}$ is the bare trap's width.  This allows us to write $R_{L0}$ as
\begin{align}\label{RLzero}
R_{L0}=
\left((N-1)\frac{2^{(d+1)/2}}{\pi^{(d-1)/2}}\frac{d(q+d)}{q}
\frac{ar_0^{2+q}}{\rho_0^D}\right)^{1/(q+d)},
\end{align}
where
\begin{align}\label{rzero}
r_0=\left(\frac{\hbar^2}{Mk}\right)^{1/(2+q)}
\end{align}
is a characteristic length associated with the bare, power-law longitudinal trapping potential.  For a harmonic longitudinal trapping potential $V_L({\bm r})=\frac12 M\omega_L^2|{\bm r}|^2$, this characteristic length reduces to the bare width of the longitudinal trap, $r_0=\sqrt{\hbar/M\omega_L}$.

% section longitudinal_thomas_fermi_approximation (end)

% ================================================================================
\section{Ground-state properties \label{sec:results}} % (fold)

Aside from shedding light on the physical behavior of lower-dimensional BECs, the main objective of a reduced-dimension, mean-field model is to provide a simpler way to estimate accurately the condensate's ground-state properties without having to solve the 3D GP equation~(\ref{3DGPE}) numerically.  Such a simplification is possible as long as the transverse spatial degrees of freedom have little effect in determining the desired quantities.  Below we assess the validity and performance of the optimal product-state approximation by deriving analytical formulas to predict various physical properties of the condensate, which we compare against numerical integration of the 3D GP equation~(\ref{3DGPE})~\cite{dion07} for quasi-1D and -2D potentials with different aspect ratios and various atom numbers.  Additionally, we benchmark the performance of our approach against the variational method described in Appendix~\ref{ap:variational_method}, which utilizes spatially entangled trial functions.

\subsection{Numerical simulations} % (fold)
\label{sub:numerical_simulations}

In the numerical results presented below, we consider highly anisotropic condensates of $^{87}$Rb atoms in the $|F = 1, m_F = -1\rangle$ hyperfine state, for which $a=100.4\,a_0$, with $a_0$ being the Bohr radius. Although our method applies to general trapping potentials, we restrict the following numerical analysis to the case of quasi-1D and -2D harmonic potentials,
\begin{equation}
\label{potential}
    V({\bm \rho},{\bm r}) = \frac{1}{2}M\!\left(\omega_T^2 |{\bm \rho}|^2 + \omega_L^2 |{\bm r}|^2\right),
\end{equation}
with different aspect ratios.  We set the longitudinal frequency $\omega_L/2\pi$ to 3.5~Hz and the transverse frequency $\omega_T/2\pi$ to three different values: 35, 175, and 350~Hz.  For these parameters, the bare trap's widths, $r_0=\sqrt{\hbar/M\omega_L}$ and $\rho_0=\sqrt{\hbar/M\omega_T}$, are such that $r_0\simeq~5.8\,\mu$m, and the aspect ratio of the bare traps, $\rho_0:r_0$, is approximately 1:3, 1:7, and 1:10, respectively.  Moreover, in Eq.~(\ref{NL}) these frequencies give a lower critical atom number $N_L\simeq 70, 15,$ and 8 for $d=1$ and $N_L\simeq 430, 190,$ and 140 for $d=2$.  From Eq.~(\ref{NT}), we find an upper critical atom number $N_T\simeq4\,500, 10\,000,$ and 14\,000 for $d=1$ and $N_T\simeq12\,000, 135\,000,$ and 380\,000 for $d=2$.   When the traps are loaded with $N = N_T$ atoms, the condensate aspect ratio, $\rho_0$:$R_{L0}$, becomes 1:16, 1:79, and 1:158 for $d=1$ and 1:12, 1:61, and 1:122 for $d=2$.  

We point out that for the traps we consider below, $\rho_0\gg a$ and thus the spherically symmetric $s$-wave scattering is not appreciably distorted by the tight transverse confinement.  If, however, $\rho_0\lesssim a$, further corrections become necessary~\cite{olshanii}.

% section numerical_simulations (end)

\subsection{Chemical potential} % (fold)
\label{sub:chemical_potential}

\begin{figure}[htbp]
	\centering
	\includegraphics[width=4.5truein]{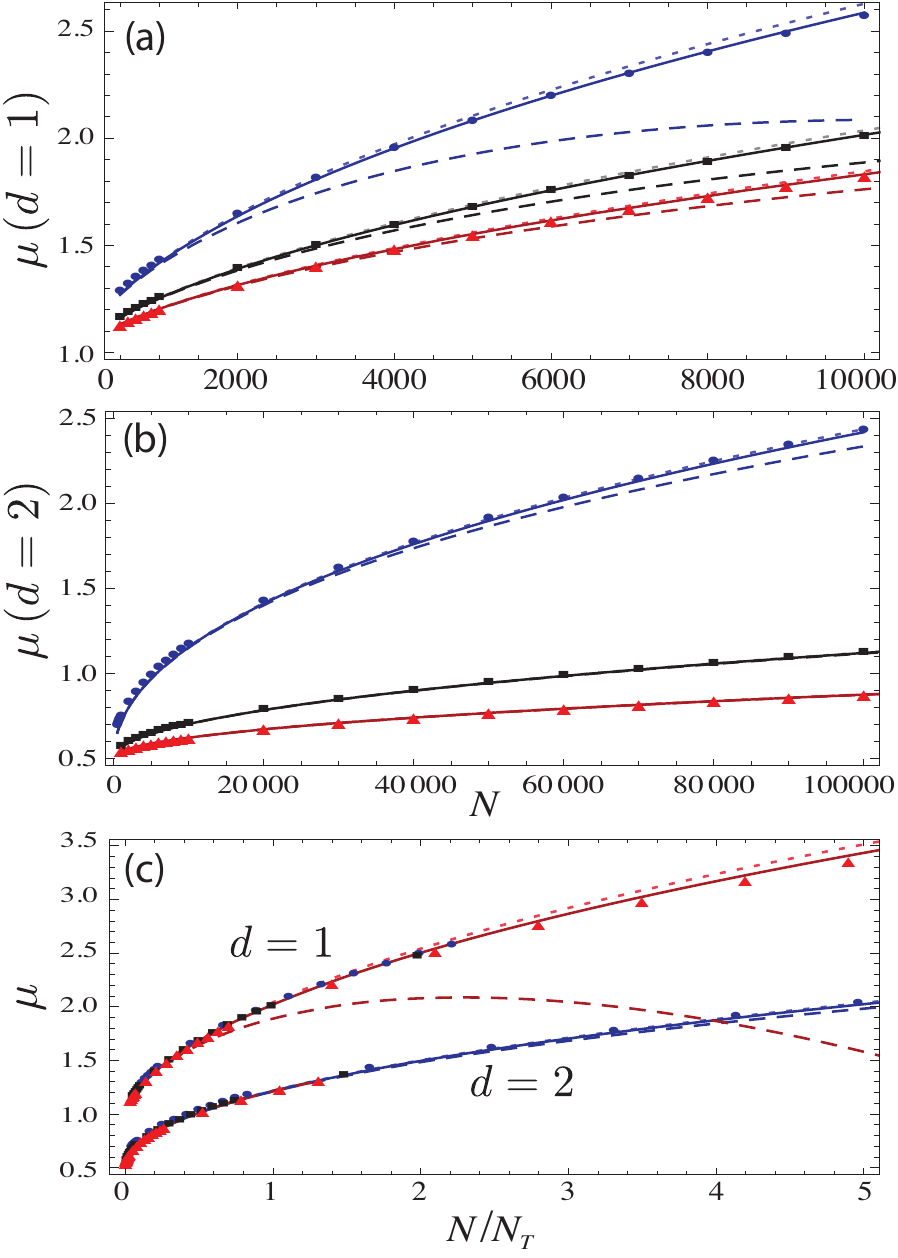}
	\caption{(Color online) Chemical potential $\mu$ in units of $\hbar\omega_T$ for different (a)~quasi-1D and (b)~-2D harmonic potentials as a function of atom number.  Points correspond to numerical integration of the 3D GP equation~(\ref{3DGPE}) for a fixed longitudinal frequency of $3.5$~Hz and transverse frequencies equal to 35~Hz (blue circles), 175~Hz (black squares), and 350~Hz (red triangles). Solid lines show the result from formula~(\ref{muL}) with $R_L$ being the exact (numerical) solution to Eq.~(\ref{RL}). Dashed lines show the first-order formula~(\ref{muL1st_harmonic}). Dotted lines correspond to formula~(\ref{muL}), but with the longitudinal TF length and radius given by Eqs.~(\ref{R1L}) and (\ref{R2L}), which are obtained via the variational method discussed in Appendix~\ref{ap:variational_method}.  The agreement between the full calculation (solid lines) and the exact 3D numerical result is indeed remarkable throughout and beyond the reduced-dimension regime.  Plot~(c) shows the chemical potential as a function of $N/N_T$, which is independent of the trap's aspect ratio.}
	\label{fig:Mu}
\end{figure}

According to our perturbation theory, the chemical potential, $\mu\simeq E_0+\mu_L$, is determined solely by the optimal product state (dominant Schmidt term) up to first-order in $\epsilon$; the transverse wave function $\chi_0$ contributes only through the zero-point energy $E_0$, whereas the nontrivial part of the chemical potential, $\mu_L$, is provided entirely by the longitudinal wave function $\phi_0$, as instructed by Eqs.~(\ref{muL}) and (\ref{RL}).  In Fig.~\ref{fig:Mu}, we plot such a prediction against the exact chemical potential given by the numerical integration of the 3D GP equation~(\ref{3DGPE}) for different quasi-1D [$d=1$, Fig.~\ref{fig:Mu}(a)] and -2D [$d=2$, Fig.~\ref{fig:Mu}(b)] harmonic potentials as a function of atom number.  We evaluate Eq.~(\ref{muL}) in two ways: first, by using the first-order formula~(\ref{muL1st}) for $q=2$, given by
\begin{align}
	\mu_L
&=\frac{1}{2} \hbar\omega_T\!\left( \frac{\rho_0R_{L0}}{r_0^2} \right)^{\!2}
\!\left( 1 - \epsilon \frac{12}{(d+4)(d+2)}
\frac{\UpsilonT \hbar\omega_T}{\eta_T^2}
\!\left( \frac{\rho_0R_{L0}}{r_0^2} \right)^{\!2}\right),
\label{muL1st_harmonic}
\end{align}
which follows from the first-order estimate of $R_L$, given by Eq.~(\ref{RL1st}); second, by using the estimate of $R_L$ given by the exact solution of Eq.~(\ref{RL}).  Here and elsewhere in this section, we use $k=\hbar\omega_T\rho_0^2/r_0^4$, $R_{L0}$ is evaluated from Eq.~(\ref{RLzeroharmonic}) or (\ref{RLzeroNT}), and $\UpsilonT\hbar\omega_T/\eta_T^2$ from Eq.~(\ref{Upsilon_harmonic}).  For both quasi-1D and -2D geometries and different aspect ratios, Fig.~\ref{fig:Mu} shows that the first-order formula~(\ref{muL1st_harmonic}) is in good agreement with the exact numerical results in the reduced-dimension regime $N\ll N_T$.  As expected, this formula breaks down as $N$ approaches $N_T$, since the normalization condition~(\ref{RL}) is not exactly satisfied.  Interestingly, such a divergence is much slower for quasi-2D than for quasi-1D traps, a fact that we discuss further in Appendix~\ref{ap:critical_atom_numbers}.  Note, however, that once we utilize the exact solution for $R_L$, thus satisfying the normalization condition, we are able to get a better estimate of the chemical potential throughout and even beyond the reduced-dimension regime, as shown in Fig.~\ref{fig:Mu}(c).  The agreement between this full calculation and the exact 3D numerical result is indeed remarkable and even better than the variational technique for both geometries and all aspect ratios that we consider.  This plot also makes clear that the different trap frequencies only change the value of $N_T$ and thus define different reduced-dimension regimes of atom numbers.  Thus, for a fixed atom number, the agreement with the exact numerical results improves with increasing aspect ratio, because $N_T$ also increases with increasing aspect ratio. However, once renormalized by $N_T$, the chemical potential becomes independent of the trap's aspect ratio.  As we discuss below, the same effect is observed for other ground-state quantities.  Finally, it is worth pointing out that our formulas hold even in the case of the modestly anisotropic trap $\omega_T/2\pi = 35$ Hz, for which $r_0/\rho_0\simeq 3$.

% subsection chemical_potential (end)

\subsection{Average density} % (fold)
\label{sub:average_density}

As the results above indicate, the chemical potential only has higher-order support on the second Schmidt term and, hence, provides no information about its importance (or about the spatial correlations) in the reduced-dimension regime. To this end, however, we calculate the condensate's average density
\begin{align}
    N\eta &= N\int d^D\!\rho\,d^dr\,|\uppsi({\bm \rho},{\bm r})|^4,\label{Neta_definition}
\end{align}
which according to our perturbation theory formally depends on the second Schmidt term to first order in $\epsilon$, since
\begin{align}
    \eta &\simeq \int d^D\!\rho\,d^dr\left[ \chi_0^4({\bm \rho})\phi_0^4({\bm r}) + 4 \epsilon \chi_0^3({\bm \rho})\phi_0^3({\bm r})\chi_1({\bm \rho})\phi_1({\bm r})\right].\label{eta}
\end{align}
Here $\eta$ is a measure of the inverse volume occupied by the ground-state wave function, which can also be thought of as the average density per atom; i.e., $N\eta$ is the density, $N|\uppsi({\bm \rho},{\bm r})|^2$, averaged over the probability density $|\uppsi({\bm \rho},{\bm r})|^2$.

From Eqs.~(\ref{chi0}), (\ref{chi1}), (\ref{phi1}) and (\ref{phi0TF}), we find to first order in $\epsilon$,
\begin{align}
    \eta &=\eta_T\tilde\eta_L + 2\epsilon\tilde g\UpsilonT\!\left( \eta_L^2 + \Delta\eta_L^2\right) +O(\epsilon^2),\label{eta2}
\end{align}
where
\begin{align}
	\tilde\eta_L&=\int_{r\le R_L}d^dr\left( \frac{\mu_L-V_L({\bm r})}{\tilde{g}\eta_T} \right)^{\!2}
    = \eta_L\!\left( \frac{R_L}{R_{L0}} \right)^{2q+d}
    \simeq \eta_L\!\left(1-\epsilon\frac{\UpsilonT}{\eta_{T}^2} \frac{3q}{q+d}k R_{L0}^q\right).\label{etaLtilde}
\end{align}
Equation~(\ref{eta2}) and the first equality in Eq.~(\ref{etaLtilde}) apply to all trapping potentials, and the second equality in Eq.~(\ref{etaLtilde}) specializes to power-law longitudinal potentials.

For harmonic trapping potentials, we use Eqs.~(\ref{etaL_harmonic}), (\ref{DeltaetaL_harmonic}), and (\ref{getaT}) or (\ref{con}) to get
\begin{align}
	N\eta &\simeq N\eta_T\eta_L\!\left(1- \epsilon\frac{24}{(d+2)(d+6)}\frac{\UpsilonT \hbar\omega_T}{\eta_T^2}\!\left( \frac{\rho_0 R_{L0}}{r_0^2} \right)^{\!2} \right).\label{Neta}
\end{align}
In the limit $\epsilon\rightarrow 0$, this reduces, as expected, to
\begin{align}
	N \eta_T\eta_L=N\!\left( \frac{1}{\sqrt{2\pi} \rho_0} \right)^D\frac{d(d+2)}{(d+4)\pi^{d-1}}\frac{1}{R_{L0}^d}.
\end{align}

\begin{figure}[htbp]
	\centering
	\includegraphics[width=4.5truein]{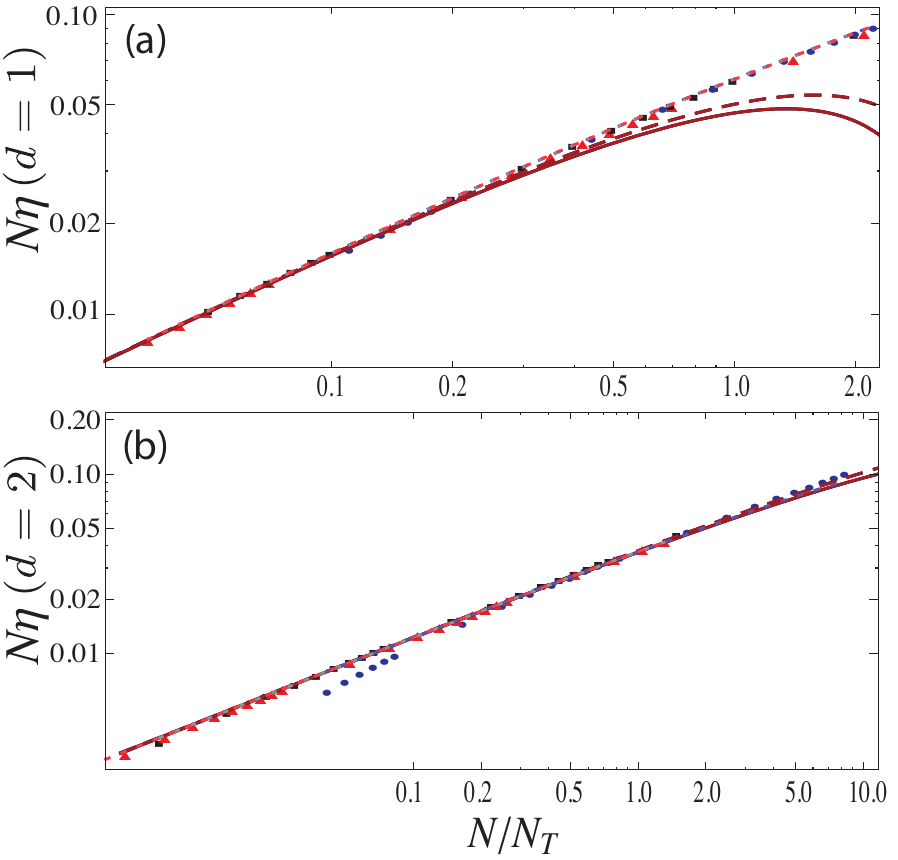}
	\caption{(Color online) Average density~(\ref{Neta_definition}) as a function of $N/N_T$ for different (a)~quasi-1D and (b)~-2D harmonic potentials in units of $(\rho_0^2 a)^{-1}$.  Points correspond to numerical integration of the 3D GP equation~(\ref{3DGPE}) for a fixed longitudinal frequency of $3.5$~Hz and transverse frequencies equal to 35~Hz (blue circles), 175~Hz (black squares), and 350~Hz (red triangles). Solid lines show the result from the first-order formula~(\ref{Neta}), which contains the contributions of the two Schmidt terms.  Dashed lines, on the other hand, show the first-order formula~(\ref{Neta0}), which is determined entirely by the dominant Schmidt term, i.e., by the optimal product approximation. Dotted lines correspond to the prediction given by direct numerical integration of the variational wave functions~(\ref{ansatz}).  The agreement between the analytical formulas~(\ref{Neta}) and (\ref{Neta0}) and the 3D numerical results is very good for $N\ll N_T$.  Moreover, the prediction~(\ref{Neta0}) provided solely by the optimal product is as good as the full estimate~(\ref{Neta}) and the variational approach, which thus shows that corrections to the optimal product are not significant in the reduced-dimension regime.}
	\label{fig:AvgDensity}
\end{figure}

In Fig.~\ref{fig:AvgDensity}, we compare the analytical formula~(\ref{Neta}) against the exact mean-field average density obtained via direct integration of the numerical solution of the 3D GP equation for (a)~$d=1$ and (b)~$d=2$.  In addition, we plot the average density predicted by the dominant Schmidt term alone, which is given by
\begin{align}
	N\eta_0
&=N\Bigl(\eta_T\tilde\eta_L+2\epsilon\tilde g\UpsilonT(\eta_L^2+3\Delta\eta_L^2)\Bigr)\nonumber\\
&=N\Bigl(\eta + 4 \epsilon \tilde g \UpsilonT\Delta\eta_L^2\Bigr)
=N\eta_T\eta_L\!\left( 1 - \epsilon \frac{4(d^2-4d-24)}{(d+2)(d+4)(d+6)}
\frac{\UpsilonT \hbar\omega_T}{\eta_T^2}\!\left( \frac{\rho_0 R_{L0}}{r_0^2} \right)^{\!2} \right),\label{Neta0}
\end{align}
as well as the prediction given by direct numerical integration of the variational wave functions~(\ref{ansatz}).  The agreement between the 3D numerical results and the estimates from the analytical formulas~(\ref{Neta}) and (\ref{Neta0}) is very good for $N\ll N_T$.  Moreover, the prediction~(\ref{Neta0}) provided only by the dominant Schmidt term, i.e., the optimal product, is as good as the full estimate~(\ref{Neta}) and the variational approach, which shows that corrections to the optimal product are not significant in the reduced-dimension regime.  For $d=1$, deviations between the different predictions only become noticeable as $N$ approaches $N_T$, which coincides with the atom number regime in which the perturbation theory breaks down.  On the other hand, the variational technique, which inherently takes into account higher-order effects, performs remarkably well for the entire atom number regime we consider.  For $d=2$, the same behavior is observed but only for much larger atom numbers ($\gtrsim 10 N_T$).  Note that in the case of $d=2$ and $\omega_T/2\pi=35$ Hz (blue circles), the 3D numerical results also deviate from all analytical predictions for $N\lesssim 0.1 N_T$.  This is, however, consistent with all the models analyzed, since for such trap parameters, the lower critical atom number is $N_L\simeq 0.1 N_T$, and, hence, the longitudinal TF approximation is expected to break down for such small atom numbers.

% subsection average_density (end)

\subsection{Spatial entanglement} % (fold)
\label{sub:spatial_entanglement}

Although $\uppsi({\bm \rho},{\bm r})$ is formally spatially entangled, our results indicate that the transverse and longitudinal co\"ordinates are effectively decoupled for $N\ll N_T$.  For this reason, one can obtain an accurate description of the condensate mean field using only the dominant Schmidt term, which gives the optimal-product approximation and, hence, the optimal reduced-dimension mean field $\phi_0({\bm r})$.  Such a conclusion can be verified by calculating the purity of the reduced density matrices~(\ref{nT}) and (\ref{nL}), which is defined as
\begin{align}\label{purity}
	\Pi &= \int d^D\!\rho\,d^D\!\rho'\, n_T({\bm \rho},{\bm \rho'})n_T({\bm \rho'},{\bm \rho})
    =\int d^d r\,d^d r'\, n_L({\bm r},{\bm r'})n_L({\bm r'},{\bm r}).
\end{align}
The purity serves as a measure of the longitudinal-transverse spatial entanglement and satisfies the bounds $1\ge\Pi\ge 0$; $\Pi=1$ indicates that $\uppsi({\bm \rho},{\bm r})$ is spatially separable, and values of $\Pi$ less than 1 signal spatial entanglement.  As $\Pi$ decreases and approaches zero, the spatial entanglement increases.

According to the perturbative Schmidt decomposition, the longitudinal and transverse density matrices are given by Eqs.~(\ref{nTschmidt}) and (\ref{nLschmidt}), which to first-order in $\epsilon$ are pure.  Indeed Eq.~(\ref{purity}), evaluated for the wave function~(\ref{psiSchmidt}), gives
\begin{align}
	\Pi = \lambda_0^2 + O(\epsilon^4) \simeq 1 - 2\epsilon^2\lambda_1.\label{purity_schmidt}
\end{align}
Note that we enforced the normalization condition, $\lambda_0 + \epsilon^2\lambda_1 = 1$, in the approximation on the right-hand side to guarantee that the purity is bounded by 1.

Using Eq.~(\ref{chi1}), we find the Schmidt coefficient $\lambda_1=\braket{\chi_1}{\chi_1}$ to be given by
\begin{align}
	\lambda_1=
\tilde g^2\Delta\eta_L^2 \sum_{n=1}^\infty \frac{\langle\xi_n|\xi_0^3\rangle^2}{(E_n-E_0)^2}
=a^2(N-1)^2\Delta\eta_L^2
\times\begin{cases}
\displaystyle{\vphantom{\frac{\pi}{4}}{\rm Li}_2(1/4)},
&\mbox{$d=1$,}\\
\displaystyle{\frac{\pi}{4}\rho_0^2\,
{}_4F_3\!\left(1,1,1,\frac{3}{2};2,2,2;\frac{1}{4}\right)},&\mbox{$d=2$.}
	\end{cases}
\end{align}
${\rm Li}_s(z)\equiv\sum_{n=1}^\infty z^n/n^s$ is the polylogarithm function and
\begin{align}
	_pF_q(\alpha_1 \dotsb \alpha_p;\beta_1 \dotsb \beta_q;z)\equiv\sum_{n=0}^\infty \frac{(\alpha_1)_n \dotsb (\alpha_p)_n}{(\beta_1)_n \dotsb (\beta_q)_n} \frac{z^n}{n!}
\end{align}
is the generalized hypergeometric function, defined in terms of the Pochhammer's symbol $(\alpha)_n=\Gamma(\alpha+n)/\Gamma(\alpha)$~\cite{gradshteyn}.

\begin{figure}[htbp]
	\centering
	\includegraphics[width=4.5truein]{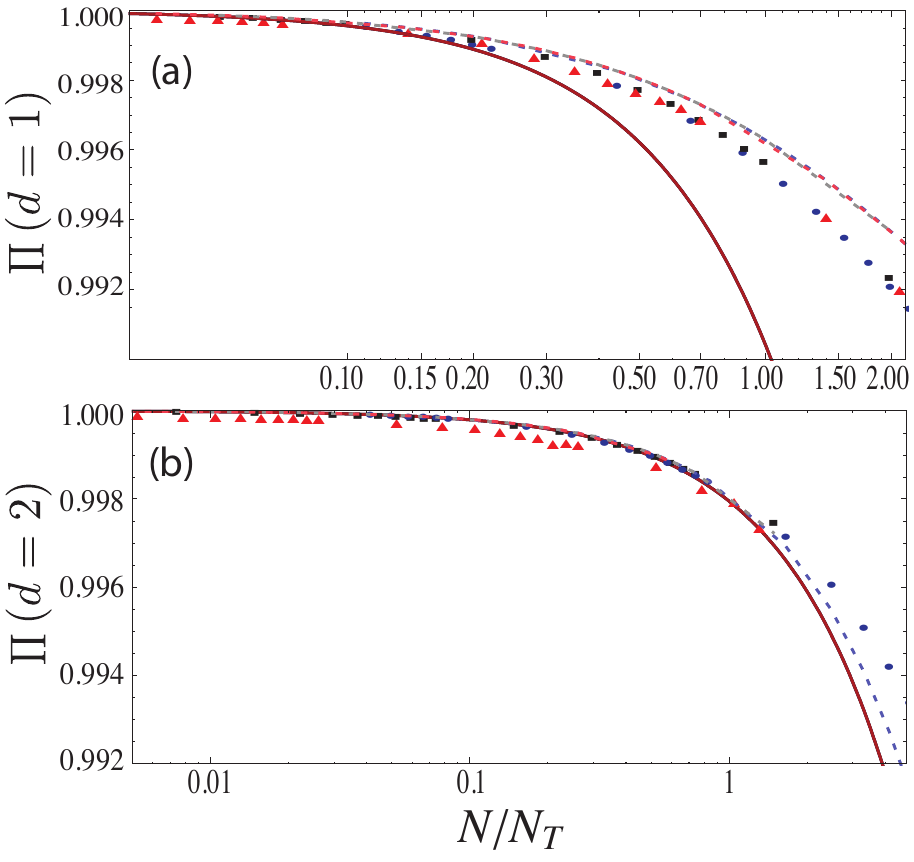}
	\caption{Purity~(\ref{purity}) of the reduced density matrices~(\ref{nT}) and (\ref{nL}) of a $^{87}$Rb condensate trapped by different (a)~quasi-1D and (b)~-2D harmonic potentials as a function of $N/N_T$.  Points correspond to the exact calculation provided by numerical integration of the 3D GP equation~(\ref{3DGPE}) for a fixed longitudinal frequency of $3.5$~Hz and transverse frequencies equal to 35~Hz (blue circles), 175~Hz (black squares), and 350~Hz (red triangles). Solid lines show the result from formula~(\ref{purity_schmidt}). Dotted lines correspond to the purity evaluated for the variational density matrices~(\ref{nL1}) and (\ref{nL2}).  Note that $\Pi\simeq 1$ for $N\ll N_T$, which reveals that $\uppsi({\bm \rho},{\bm r})$ is effectively spatially separable in the reduced-dimension regime.  As $N$ approaches $N_T$, the purity of the mean field decreases and the longitudinal-transverse entanglement can no longer be neglected. This change in the structure of the BEC wave function indicates the onset of the dimensional crossover from a lower dimensional to a 3D regime.}
	\label{fig:Purity}
\end{figure}

Figure~\ref{fig:Purity} shows the exact purity~(\ref{purity}) calculated for the numerical solution of the 3D GP equation for different (a)~quasi-1D and (b)~-2D harmonic potentials as a function of $N/N_T$, as well as its prediction given by Eq.~(\ref{purity_schmidt}) and the variational method~\cite{salasnich}, for which the (longitudinal) reduced density matrices are given by Eqs.~(\ref{nL1}) and (\ref{nL2}).  It thus become clear that $\Pi\simeq 1$ for $N\ll N_T$. Therefore, for all quasi-1D and -2D geometries we consider, the mean field $\uppsi({\bm \rho},{\bm r})$ is indeed very close to a spatially separable state throughout the reduced-dimension regime, as is assumed by our perturbation theory.    Surprisingly, the variational ansatz~(\ref{ansatz}), which is obviously not spatially separable, turns out to contain the least amount of spatial entanglement in the reduced-dimension regime among the analytical wave functions we analyze.  As $N$ approaches $N_T$, the condensate mean field crosses over from a product state to a spatially entangled regime.  Such a change in the structure of the BEC wave function signals the dimensional crossover from lower dimensionality to the full 3D regime.

% subsection spatial_entanglement (end)

% section results (end)

% ================================================================================

\section{Conclusion} % (fold)
\label{sec:conclusion}

We have studied the reduced dimensionality of BECs in nonuniform, highly anisotropic potentials in relation to the entanglement between its longitudinal and transverse spatial degrees of freedom.  Within the mean-field approximation, we have found that the condensate's reduced dimensionality is fundamentally related to a regime of near-absence of spatial entanglement.  This result can be appreciated in terms of the perturbative Schmidt decomposition of the condensate mean field~\cite{tacla11}, which not only characterizes the spatial entanglement, but also provides the optimal product-state approximation to the condensate mean field for arbitrary quasi-1D and -2D trapping potentials of the form $V_T({\bm \rho})+V_L({\bm r})$.

By taking advantage of the longitudinal TF approximation, we obtained a fully analytical estimate of the 3D mean-field solution, which we used to derive algebraic formulas for the condensate's ground-state quantities, such as the chemical potential and average density.  We have assessed the performance of the optimal product-state approximation and of the full decomposition for quasi-1D and -2D harmonic potentials for various atom numbers and aspect ratios, which we verified to be accurate in the reduced-dimension regime.  In such a representation, we have shown that one can safely ignore the nearly uncoupled spatial degrees of freedom.  The dominant longitudinal Schmidt function, $\phi_0({\bm r})$, thus corresponds to the optimal reduced-dimension, pure-state description of the 3D condensate mean field.

The good agreement with the exact 3D numerics, even for relatively large atom numbers, motivates us to study the dimensional crossover in light of the spatial entanglement beyond the perturbative regime, which we will present elsewhere.  In fact, tying the dimensionality of quantum systems to the framework of spatial entanglement may open up a broad, new perspective for investigating the physics of lower-dimensional systems in a wide variety of scenarios and trapping potentials.

% section conclusion (end)

\acknowledgments
This work was supported in part by the National Science Foundation
Grants Nos.~PHY-0903953, PHY-1005540, PHY-1212445, and PHY-13145763.

% ================================================================================
\appendix

\section{Validity of the analytical solution. Critical atom numbers} % (fold)
\label{ap:critical_atom_numbers}

Following previous work~\cite{boixo09,tacla10}, we define below two critical atom numbers, $N_{L}$ and $N_{T}$, that respectively characterize the validity of the longitudinal TF approximation and of the Schmidt perturbation theory.  We define these numbers generally for any trapping potentials, but evaluate them explicitly only for harmonic potentials in both the transverse and longitudinal directions.

The longitudinal TF approximation is valid as long as the scattering interaction is much larger than the longitudinal kinetic energy.  We thus define the lower critical atom number, $N_L$, as the atom number at which the mean-field scattering energy is as large as the longitudinal kinetic energy, i.e.,
\begin{align}
	\frac{g (N_L-1)}{2} \int d^D\!\rho\,d^dr\,|\uppsi({\bm \rho},{\bm r})|^4 = \frac{\hbar^{2}}{2M} \int d^D\!\rho\,d^dr\,|\nabla_L \uppsi({\bm \rho},{\bm r})|^2.\label{eqNL}
\end{align}
For $N\ll N_L$, the scattering term in~(\ref{3DGPE}) is negligible compared to the longitudinal energy scale, and $\uppsi({\bm\rho},{\bm r})$ is the product ground-state wave function of the Schr\"odinger equation, given by
\begin{equation}
    \uppsi({\bm\rho},{\bm r}) \simeq \xi_0({\bm \rho})\varphi_0({\bm r}),
\end{equation}
where $\xi_0$($\varphi_0$) is the ground-state wave function of $H_{T(L)}$.  Within this approximation and for the particular case of harmonic trapping potentials, for which $\eta_T$ is given by Eq.~(\ref{etaT_harmonic}) and similarly $\int d^d r\,|\varphi_0|^4=1/(2\pi)^{d/2}r_0^d$, the right-hand side of Eq.~(\ref{eqNL}) becomes
\begin{align}\label{rhs}
\frac{\hbar^2}{M}\frac{d}{4r_0^2},
\end{align}
and the left-hand side becomes
\begin{align}
\frac{g(N_L-1)\eta_T}{2}\int d^d r\,|\varphi_0(\bm r)|^4=
\frac{\hbar^2}{M}\frac{(N_L-1)a}{\sqrt{2\pi}\rho_0^D r_0^d},
\end{align}
giving
\begin{align}
	\label{NL}
	N_L = 1+d\sqrt{\frac{\pi}{2}}\frac{r_0}{2a}\!\left(\frac{\rho_0}{r_0}\right)^D.
\end{align}

To define the upper critical atom number, $N_T$, we note that the perturbation theory should hold as long as the scattering interaction is much smaller than the transverse energy scales. Thus, we define $N_T$ as the atom number at which the mean-field scattering energy becomes as large as the transverse kinetic energy, i.e.,
\begin{align}
	\frac{g (N_T-1)}{2} \int d^D\!\rho\,d^dr\, |\uppsi({\bm \rho},{\bm r})|^4 = \frac{\hbar^{2}}{2M}\int d^D\!\rho\,d^dr\,|\nabla_T \uppsi({\bm \rho},{\bm r})|^2.\label{eqNT}
\end{align}
In this way, $N_T$ characterizes when the condensate wave function starts to spread in the transverse dimensions.  To estimate $N_T$, we note that for $N_L\ll N\ll N_T$, the longitudinal kinetic energy is negligible, and $\uppsi({\bm\rho},{\bm r})$ can be approximated by the product of the transverse ground-state wave function $\xi_0(\bm\rho)$ and the longitudinal TF solution~(\ref{phi00TF_harmonic}), which gives for the right-hand side of Eq.~(\ref{eqNT}),
\begin{align}
\frac{\hbar^2}{M}\frac{D}{4\rho_0^2}.
\end{align}
and for the left-hand side,
\begin{align}
\frac{g (N_T-1)\eta_T\eta_L}{2}
=\frac{\hbar^2}{M}(N_T-1)\frac{2^{(d-1)/2}}{\pi^{(d-1)/2}}
\frac{d(d+2)}{d+4}\frac{a}{\rho_0^D\hat R_{L0}^d},
\end{align}
where we use Eq.~(\ref{etaT_harmonic}) for $\eta_T$ and Eq.~(\ref{etaL_harmonic}), evaluated at $N=N_T$, for $\eta_L$.  Hence, $\hat R_{L0}$ is the longitudinal length scale of Eq.~(\ref{RLzeroharmonic}) evaluated at $N=N_T$.

Putting this together, we have
\begin{equation}
\hat R_{L0}^d=4(N_T-1)\frac{2^{(d-1)/2}}{\pi^{(d-1)/2}}\frac{d(d+2)}{D(d+4)}a\rho_0^{d-1}.
\label{eq:NT}
\end{equation}
Combining this result for $\hat R_{L0}$ with that in Eq.~(\ref{RLzeroharmonic}) gives
\begin{align}
\hat R_{L0}= \frac{\sqrt{D(d+4)}}{2}\frac{r_0}{\rho_0}r_0,\label{RL0_NT},
\end{align}
and this leads to
\begin{align}
	\label{NT}
	N_T = 1 +\frac{\pi^{(d-1)/2}}{8^{(d+1)/2}}\frac{[D(d+4)]^{d/2+1}}{d(d+2)}
\frac{\rho_0}{a}\!\left( \frac{r_0}{\rho_0} \right)^{2d}
\end{align}
and
\begin{align}\label{RLzeroNT}
R_{L0}=\left(\frac{N-1}{N_T-1}\right)^{1/(d+2)}\hat R_{L0}
=\left(\frac{N-1}{N_T-1}\right)^{1/(d+2)}\frac{\sqrt{D(d+4)}}{2}\frac{r_0}{\rho_0}r_0.
\end{align}

The upper critical atom number plays an important role in determining the physical version of the dimensionless expansion parameter $\epsilon$ for our perturbation theory.  By inspection of, say, Eq.~(\ref{muL1st}) for the longitudinal chemical potential, this parameter is given by
\begin{align}
\frac{3\UpsilonT}{\eta_T^2}kR_{L0}^q
\sim
\epsilon\equiv
\frac{3\UpsilonT\hbar\omega_T}{\eta_T^2}\!\left(\frac{N-1}{N_T-1}\right)^{2/(d+2)}
=\left(\frac{N-1}{N_T-1}\right)^{2/(d+2)}\times
\begin{cases}
\displaystyle{\vphantom{\Bigg(\biggr)}\frac{3}{2}\ln\frac{4}{3}\simeq0.43\;,}&\mbox{$d=1$ (cigar),}\\
\displaystyle{\vphantom{\Bigg(\biggr)}3\ln(8-4\sqrt{3})\simeq 0.21\;,}&\mbox{$d=2$ (pancake).}
\end{cases}.
\label{epsilon}
\end{align}
The expression on the left is valid for arbitrary longitudinal power-law potentials and arbitrary transverse potentials, and the two on the right, defined to be $\epsilon$, specialize to harmonic longitudinal and transverse potentials.  The final expression uses the explicit expression for $\Upsilon_T$ for harmonic traps given in Eq.~(\ref{Upsilon_harmonic}).  In all these expressions, we omit $d$-dependent prefactors on the grounds that we are interested in the scaling of $\epsilon$ with $N$ and that these prefactors vary from one quantity to the next.   Note that atomic and trap properties enter into $\epsilon$ only by setting the value of the upper critical atom number $N_T$, and for this reason, they only define the reduced-dimension regime of atom numbers.  Both the pre-factor and the scaling, $(N/N_T)^{1/2}$ for $d=2$ and $(N/N_T)^{2/3}$ for $d=1$, indicate that the perturbation theory should work better for pancakes, a fact that we have also verified from the results of Sec.~\ref{sec:results}.
% section critical_atom_numbers (end)

\section{Harmonic potentials} % (fold)
\label{ap:harmonic_potentials}

In the case of harmonic trapping potentials, the transverse ground-state wave function is the Gaussian
\begin{equation}
\label{gaussian}
	\xi_{0}({\bm \rho}) = \frac{e^{-\rho^2/2\rho_0^2}}{(\pi\rho_0^2)^{D/4}}\;,
\end{equation}
with $\rho_0 = \sqrt{\hbar/M\omega_T}$, whereas $\phi_{00}$ can be written in the longitudinal TF approximation as
\begin{align}
\label{phi00TF_harmonic}
	\phi_{00}({\bm r})=\sqrt{\frac{d(d+2)}{4\pi^{d-1}R_{L0}^d}\!\left(1-\frac{|{\bm r}|^2}{{R_{L0}^2}}\right)}.
\end{align}
Thus it is easy to obtain the relations
\begin{align}
	\eta_T &= \frac{1}{(2\pi)^{D/2}\rho_0^D},\label{etaT_harmonic}\\
	\eta_L & =\frac{d(d+2)}{(d+4)\pi^{d-1}}\frac{1}{R_{L0}^d}
    =\frac{1}{(2\pi)^{(d-1)/2}(d+4)(N-1)}\frac{\rho_0^D R_{L0}^2}{ar_0^4},\label{etaL_harmonic}\\
	\Delta\eta_L&=\sqrt{\frac{d}{2 (d+6)}}\eta_L.\label{DeltaetaL_harmonic}
\end{align}
In these expressions, the length $R_{L0}$ is the harmonic ($q=2$) specialization of Eq.~(\ref{RLzero}):
\begin{align}\label{RLzeroharmonic}
R_{L0}=
\left((N-1)\frac{2^{(d-1)/2}d(d+2)}{\pi^{(d-1)/2}}\frac{ar_0^4}{\rho_0^D}\right)^{1/(d+2)}.
\end{align}

The coupling constants in Eq.~(\ref{effGPE_TFapprox}) can also be calculated explicitly. For a pancake ($D=1$), we have
\begin{equation}\label{overlap1D}
\langle \xi_{n} | \xi_0^3 \rangle=
\begin{cases}
\displaystyle{\frac{(-1)^{n/2}}{\sqrt{\pi\, n!}}\eta_T \Gamma\!\left(\frac{n + 1}{2}\right)}\;,
&\mbox{$n$ even,}\\
0\;,&\mbox{$n$ odd.}
\end{cases}
\end{equation}
For a cigar ($D=2$), if we use polar co\"ordinates for the transverse eigenfunctions, they take the form $\xi_{n_r m}(\rho,\varphi)$, with $n_{r}$ and $m$ being the radial and azimuthal quantum numbers and with the eigenenergies given by $E_{n_r m}=\hbar\omega_T(2n_r+|m|+1)$.  Then we find that
\begin{equation}
\langle \xi_{n_{r}m} | \xi_{00}^3 \rangle=2^{-n_{r}}\eta_T\delta_{m0}\;.\label{overlap2D}
\end{equation}
It follows from Eq.~(\ref{UpsilonT}) that $\UpsilonT$ is given by
\begin{equation}
\label{Upsilon_harmonic}
\UpsilonT=\frac{\eta_T^2}{\hbar\omega_T}\times
\begin{cases}
\displaystyle{\frac{1}{2}\ln\frac{4}{3}\;,}&\mbox{$d = 1$ (cigar),}\\
\displaystyle{\vphantom{\frac12}\ln(8-4\sqrt{3})\;,}&\mbox{$d=2$ (pancake).}\\
\end{cases}
\end{equation}
As a result, the coupling constants that appear in various equations of our perturbation theory are
\begin{align}\label{getaT}
	g \eta_T = 2\hbar\omega_T a(2\pi)^{(d-1)/2}\rho_0^{d-1}=2\hbar\omega_T a\times
	\begin{cases}
\displaystyle{1\;,}&\mbox{$d=1$ (cigar),}\\
\displaystyle{\sqrt{2\pi}\rho_0\;,}&\mbox{$d = 2$ (pancake),}
	\end{cases}
\end{align}
and
\begin{align}
	3g^2\UpsilonT = 6\hbar\omega_T a^2\times
	\begin{cases}
\displaystyle{\vphantom{\Big(}\ln(4/3)\;,}&\mbox{$d=1$ (cigar),}\\
\displaystyle{\vphantom{Big(}4\pi\rho_0^2\ln(8-4\sqrt{3})\;,}&\mbox{$d = 2$ (pancake).}
	\end{cases}
\end{align}
One other useful combination is
\begin{equation}\label{con}
\frac{\tilde g\eta_T}{\hbar\omega_T}\eta_L=\frac{2}{d+4}\!\left(\frac{\rho_0R_{L0}}{r_0^2}\right)^{\!2}.
\end{equation}
% section harmonic_potentials (end)

\section{Variational Gaussian method} % (fold)
\label{ap:variational_method}

Assuming a tight, transverse harmonic trapping potential $V_T({\bm \rho})=m\omega_T|{\bm \rho}|^2/2$ and a loose, $d$-dimensional longitudinal potential $V_L({\bm r})$, the variational approach by Salasnich \emph{et al.}~\cite{salasnich} proposes a (unity normalized) trial wave function of the form
\begin{align}
	\psi_d({\bm \rho},{\bm r}) = \frac{e^{-|{\bm \rho}|^2/2\sigma_d({\bm r})^2}}{\pi^{1/(2d)}\sigma_d^{1/d}({\bm r})}f_d({\bm r}),\label{ansatz}
\end{align}
where the variational functions $\sigma_d({\bm r})$ and $f_d({\bm r})$ are obtained by using Eq.~(\ref{ansatz}) to minimize the mean-field action. The \emph{Ansatz}~(\ref{ansatz}) is not spatially separable, as it describes the transverse spatial co\"ordinate(s) by a Gaussian wave function whose width $\sigma_d({\bm r})$ varies longitudinally.

Note that $f_d({\bm r})$ provides a direct estimate of the longitudinal marginal distribution,
\begin{align}
	\int d^D\!\rho\,|\psi_a({\bm \rho},{\bm r})|^2=|f_d({\bm r})|^2,
\end{align}
which corresponds to the diagonal part of the longitudinal reduced density matrix $n_{Ld}({\bm r},{\bm r'})$, given by
\begin{align}
	n_{L1}(z,z')&=\frac{2 f_1(z) \sigma_1(z) f_1(z') \sigma_1(z')}{\sigma_1^2(z)+\sigma_1^2(z')} \quad (d=1),\label{nL1}\\
	n_{L2}(\varrho,\varrho')&= \frac{\sqrt{2 \sigma_2(\varrho)\sigma_2(\varrho')}f_2(\varrho) f_2(\varrho')}{\sqrt{ \sigma_2^2(\varrho)+\sigma_2^2(\varrho')}} \quad (d=2),\label{nL2}
\end{align}
where in the case of cylindrically symmetric potentials, $z$ corresponds to the axial direction and $\varrho$ to the radial.  That is, for a quasi-1D trap ($d=1$), $z$ labels the longitudinal co\"ordinate and $\varrho$ the transverse radial co\"ordinate; for $d=2$, the labels are reversed, with $z$ labeling the transverse co\"ordinate and $\varrho$ the longitudinal radial co\"ordinate.  Note that the radial co\"ordinate $\varrho$ in Eq.~(\ref{nL2}) should not be confused with $\rho$, which labels the tight, transverse direction throughout this paper, regardless of the trap geometry.

Thus, one finds for quasi-1D ($d=1$) potentials
\begin{align}
	\sigma_1^2 &= \rho_0^2 \sqrt{1+2\tilde a|f_1|^2},\label{sigma1d}\\
	\mu_1 f_1 &= \left[H_L +\frac{\tilde g}{2\pi\rho_0^2}\frac{|f_1|^2}{\sqrt{1+2\tilde a|f_1|^2}}+ \frac{\hbar\omega_T}{2}\!\left( \frac{1}{\sqrt{1+2\tilde a|f_1|^2}} +\sqrt{1+2\tilde a|f_1|^2} \right) \right] f_1, \label{NPSE1d}
\end{align}
where $\rho_0=\sqrt{\hbar/m\omega_T} $, $\mu_1$ is the method's estimate of the chemical potential, and $\tilde a= a(N-1)$.  Within the longitudinal Thomas-Fermi approximation, Eq.~(\ref{NPSE1d}) assumes the form
\begin{align}
 \frac{\mu_1-V_L}{\hbar\omega_T} = \frac{1+3\tilde a|f_1|^2}{\sqrt{1+2\tilde a|f_1|^2}}.\label{NPSE1dTF}
\end{align}
which can be further simplified in the reduced-dimension regime
\begin{align}
	% \sigma_1^2 &\simeq \rho_0^2 (1+\tilde a|f_1|^2),\\
	\frac{\mu_1-V_L}{\hbar\omega_T}&\simeq 1+ 2\tilde a|f_1|^2 - \frac{3}{2}a^2 N^2|f_1|^4,\label{f1quintic}
\end{align}
where we assumed that $\tilde a|f_1|^2\ll 1$.  Interestingly, within this regime, the nonpolynomial Schr\"odinger equation~(\ref{NPSE1d}) [or equivalently Eq.~(\ref{NPSE1dTF})] also provides a reduced-dimension GP equation for the longitudinal degrees of freedom with an additional three-body, attractive interaction. Note, however, that the strength of the three-body coupling is different from that of Eq~(\ref{phi0}).  It is easy to show that Eq.~(\ref{f1quintic}) has the following perturbative physical solution
\begin{align}
	|f_1|^2=\frac{\mu_{1L}-V_L}{2\tilde a\hbar\omega_T}+ \frac{3\tilde a}{4}\!\left( \frac{\mu_{1L}-V_L}{2\tilde a\hbar\omega_T} \right)^{\!2},
\end{align}
where $\mu_{1L}=\mu_1-\hbar\omega_T$ is found from the normalization of $f_1$. In terms of the Thomas-Fermi length $R_{1L}\equiv \sqrt{2\mu_{1L}/(m \omega_L)}$, this requires solving the following equation
\begin{align}
	1=\frac{m\omega_L^2}{3 \tilde a \hbar\omega_T}R_{1L}^3 +\frac{m^2 \omega_L^4}{20 \tilde a \hbar^2 \omega_T^2}R_{1L}^5.\label{R1L}
\end{align}

For quasi-2D traps, $\sigma_2$ and $f_2$ are defined by the equations~\cite{note},
\begin{align}
	&\frac{\hbar^2}{2m\sigma_2^2}-\frac{1}{2} m\omega_T^2\sigma_2^2 +  \frac{\tilde g}{2\sqrt{2\pi}\sigma_2}|f_2|^2  =0,\label{sigma2d}\\
	&\mu_2 f_2 = \left[ H_L + \frac{\tilde g}{\sqrt{2\pi}\sigma_2}|f_2|^2 + \frac{\hbar^2}{4m\sigma_2^2} +\frac{1}{4}m\omega_T^2\sigma_2^2  \right]f_2,\label{NPSE2d}
\end{align}
where $\mu_2$ is the (quasi-2D) chemical potential.  Equation~(\ref{sigma2d}) is a quartic equation in $\sigma_2$ that can solved analytically~\cite{salasnich}.  However, in the reduced-dimension regime, it is sufficient to use the perturbative solution
\begin{align}
	\sigma_2 \simeq \rho_0\!\left( 1+\sqrt{\frac{\pi }{2}} \rho_0 \tilde a |f_2|^2\right),\label{sigma2}
\end{align}
where we used that $\rho_0 \tilde a |f_2|^2\ll 1$.  Thus, by using Eq.~(\ref{sigma2}) together with the longitudinal Thomas-Fermi approximation, Eq.~(\ref{NPSE2d}) can be simplified to
\begin{align}
 	\frac{\mu_2 - V_L}{\hbar\omega_T} \simeq \frac{1}{2} + 2\sqrt{2\pi}\rho_0 \tilde a |f_2|^2 - \frac{3\pi}{2}\rho_0^2\tilde a^2|f_2|^4,
\end{align}
which has the perturbative solution
\begin{align}
	|f_2|^2=\frac{\mu_{2L}-V_L}{2\sqrt{2\pi}\rho_0 \tilde a\hbar\omega_T}+ \frac{3\sqrt{\pi}\rho_0 \tilde a}{4\sqrt{2}}\!\left( \frac{\mu_{2L}-V_L}{2\sqrt{2\pi} \rho_0 \tilde a\hbar\omega_T} \right)^{\!2},
\end{align}
where the longitudinal part of the chemical potential $\mu_{2L}=\mu_2-\hbar\omega_T/2\equiv m\omega_L^2R_{2L}^2/2$ is determined from the normalization equation
\begin{align}
	1=\frac{m \sqrt{\pi} \omega_L^2}{8\sqrt{2} \tilde a \rho_0 \hbar\omega_T}R_{2L}^4+\frac{m^2 \sqrt{\pi} \omega_L^4}{128\sqrt{2} \tilde a \rho_0 \hbar^2 \omega_T^2}R_{2L}^6.\label{R2L}
\end{align}

% section variational_method (end)

% ================================================================================


\begin{thebibliography}{20}
\bibitem{note}
In~\cite{salasnich}, the last two terms of Eq.~(\ref{NPSE2d}) for $f_2$, which come from integrating out the transverse Gaussian, are missing a factor of 1/2.
\end{thebibliography}
\end{document}